\pdfoutput=1
\RequirePackage{ifpdf}
\ifpdf 
\documentclass[pdftex]{sigma}
\else
\documentclass{sigma}
\fi

\usepackage[all]{xy}

\numberwithin{equation}{section}

\begin{document}

\newcommand{\arXivNumber}{1402.2158}

\allowdisplaybreaks

\renewcommand{\thefootnote}{$\star$}

\renewcommand{\PaperNumber}{073}

\FirstPageHeading

\ShortArticleName{Big Bang, Blowup, and Modular Curves: Algebraic Geometry in Cosmology}

\ArticleName{Big Bang, Blowup, and Modular Curves:\\
Algebraic Geometry in Cosmology\footnote{This paper is a~contribution to the Special Issue on Noncommutative Geometry
and Quantum Groups in honor of Marc A.~Rief\/fel.
The full collection is available at
\href{http://www.emis.de/journals/SIGMA/Rieffel.html}{http://www.emis.de/journals/SIGMA/Rieffel.html}}}

\Author{Yuri I.~MANIN~$^\dag$ and Matilde MARCOLLI~$^\ddag$}

\AuthorNameForHeading{Yu.I.~Manin and M.~Marcolli}

\Address{$^\dag$~Max-Planck-Institut f\"ur Mathematik, Bonn, Germany}
\EmailD{\href{mailto:manin@mpim-bonn.mpg.de}{manin@mpim-bonn.mpg.de}}

\Address{$^\ddag$~California Institute of Technology, Pasadena, USA}
\EmailD{\href{mailto:matilde@caltech.edu}{matilde@caltech.edu}}

\ArticleDates{Received March 01, 2014, in f\/inal form June 27, 2014; Published online July 09, 2014}

\Abstract{We introduce some algebraic geometric models in cosmology related to the ``boundaries'' of space-time: Big
Bang, Mixmaster Universe, Penrose's crossovers between aeons.
We suggest to model the kinematics of Big Bang using the algebraic geometric (or analytic) blow up of a~point~$x$.
This creates a~boundary which consists of the projective space of tangent directions to~$x$ and possibly of the light
cone of~$x$.
We argue that time on the boundary undergoes the Wick rotation and becomes purely imaginary.
The Mixmaster (Bianchi~IX) model of the early history of the universe is neatly explained in this picture by postulating
that the reverse Wick rotation follows a~hyperbolic geodesic connecting imaginary time axis to the real one.
Penrose's idea to see the Big Bang as a~sign of crossover from ``the end of previous aeon'' of the expanding and cooling
Universe to the ``beginning of the next aeon'' is interpreted as an identif\/ication of a~natural boundary of Minkowski
space at inf\/inity with the Big Bang boundary.}


\Classification{85A40; 14N05; 14G35}

\renewcommand{\thefootnote}{\arabic{footnote}} \setcounter{footnote}{0}

\section{Introduction}

\subsection{Future and past boundaries of space-times}

The current observable domain of our expanding Universe is almost f\/lat.
Hence we assume that its good model is Minkowski's space-time.
Therefore, its natural {\it future boundary} is modelled by (a domain in) the inf\/inite three-dimensional hyperplane
compactifying Minkowski 4-space to the projective 4-space.

The Big Bang model of the beginning of our Universe postulates the special role of a~certain ``time zero'' point in it,
and we will argue that a~natural {\it past boundary} is related to the algebraic geometric blow-up of this point.

We start with explaining the relevant geometry in more detail.
Below $\mathbb{R}$ (resp.~$\mathbb{C}$) always denotes the f\/ield of real (reps.~complex) numbers.
It is essential to keep track of the complex/algebraic geometry of our constructions and identify classical space-times
as real points corresponding to certain real structures, as in Penrose's twistor program, description of instantons etc.
Our main new contribution in this respect will be a~description of cosmological time in the early Universe in terms of
the geodesic f\/low on a~modular curve.

Let $M^4$ be a~4-dimensional real linear space endowed with a~non-degenerate symmetric quadratic form $q: S^2(M^4)\to
\mathbb{R}$ of signature $(1,3)$ (that is, $+ ---$).
The associated af\/f\/ine space $\mathcal{M}^4$ (essentially, $M^4$ without marked zero point) with metric induced by~$q$ is
the standard model of Minkowski space-time (in which a~time-orientation is not yet chosen explicitly).
By construction, $M^4$ acts on $\mathcal{M}^4$ by shifts (and therefore it acts also on various subsets, e.g.,
af\/f\/ine subspaces of $\mathcal{M}^4$).
If one subset is obtained by shift from another subset, we say that they are {\it parallel.} Two light cones with
possibly dif\/ferent vertices in $\mathcal{M}^4$ are parallel in this sense.

Consider now two dif\/ferent compactif\/ications of $\mathcal{M}^4$: $\overline{\mathcal{M}}^4_p$ and
$\overline{\mathcal{M}}^4_q$.

{\sloppy {\em $\overline{\mathcal{M}}^4_p$ and the future boundary.} By def\/inition, $\overline{\mathcal{M}}^4_p$ is the
4-dimensional real projective spa\-ce~${\mathbb{P}}^4(\mathbb{R})$ which consists of $\mathcal{M}^4$ and points at
inf\/inity: one point at inf\/inity corresponds to the full set of pairwise parallel lines in~$\mathcal{M}^4$.
Thus, in~$\overline{\mathcal{M}}^4_p$ the Minkowski space-time $\mathcal{M}^4$ is compactif\/ied by a~3-dimensional real
projective space $\mathbb{P}^3(\mathbb{R})$.
This boundary~$\mathbb{P}^3(\mathbb{R})_{\infty}$ is endowed with an important additional structure, namely, {\it an
embedded $2$-dimensional sphere~$S^2$} which is the common base at inf\/inity of all light cones of~$\mathcal{M}^4$.

}

{\em $\overline{\mathcal{M}}^4_q$ and the past boundary.} We start with recalling that if~$X$ is a~smooth algebraic or
analytic variety and $Y\subset X$ is a~smooth closed subvariety, one can construct a~morphism ${\rm bl}_Y: \widetilde{X} \to
X$, in which $\widetilde{X}$ is another smooth variety, ${\rm bl}_Y$ restricted to the complement $\widetilde{X}\setminus
({\rm bl}_Y)^{-1}(Y)$ def\/ines its isomorphism with the initial complement $X\setminus Y$, whereas $({\rm bl}_Y)^{-1}(Y)$ is the
projectivized normal bundle to~$Y$ in~$X$, which ${\rm bl}_Y$ projects to its base~$Y$.

In particular, if $\dim X=n$ and~$Y$ is a~point $x\in X$, then blowing it up, we get a~divi\-sor~$\mathbb{P}^{n-1}$ which
``squeezes into~$X$'' replacing the former~$x$.
If we assume that~$X$ is endowed with a~conformal class of metrics, then in the tangent space to~$x$ we have
a~canonically def\/ined null cone, whereas in the projectivized tangent space embedded in the blow up it produces the
``base'' of the null cone, the local sky of the observer located at the point~$x$ (see remark at p.~256 of~\cite{Pe1}
about the dif\/ference between the {\it light cone} which is a~global object and the {\it null cone} which we invoked
above).

This interpretation forms an essential part of the motivation for our constructions.

We will now describe $\overline{\mathcal{M}}^4_q$.
By def\/inition, it is a~smooth real quadric hypersurface $Q^4$ in a~f\/ive-dimensional projective space
$\mathbb{P}^5(\mathbb{R})$, whose equation in homogeneous coordinates is given by a~quadratic form of signature $(3,3)$.
For any point $x\in Q^4$, one can construct the linear projective 4-dimensional subspace $\mathbb{P}^4_x$ in
$\mathbb{P}^5(\mathbb{R})$ which is tangent to $Q^4$ at~$x$.
Then the intersection $\overline{L}_x:=Q^4\cap \mathbb{P}^4_x$ is isomorphic to any compactif\/ied light cone in
$\overline{\mathcal{M}}^4_p$ above.
Fix~$x$ and consider the complement $Q^4\setminus L_x$.
In this complement, through each point~$y$ there passes the (uncompactif\/ied) light cone $L_y:=\overline{L}_y\setminus
(\overline{L}_y\cap \overline{L}_x)$.
One can identify $Q^4\setminus L_x$ with af\/f\/ine Minkovski space $\mathcal{M}^4$ by projecting $Q^4\setminus L_x$
from~$x$ into any suf\/f\/iciently general hyperplane $\mathbb{P}^4(\mathbb{R})\subset \mathbb{P}^5(\mathbb{R})$.

More precisely, the same projection can be completed to a~diagram of birational morphisms (restricted to real points of
algebraic varieties def\/ined over $\mathbb{R}$)
\begin{gather}
\label{(1.1)}
\begin{split}
& \xymatrix{C\ar[r]^{{\rm bl}_x} \ar[d]^{{\rm bl}_{S^2}} & Q^4\\
 \mathbb{P}^4(\mathbb{R}) }
\end{split}
\end{gather}
Namely,~$C$ is obtained from $Q^4=\overline{\mathcal{M}}^4_q$ by blowing up the point $x\in Q^4$, and the same~$C$ is
obtained from $\mathbb{P}^4 (\mathbb{R})= \overline{\mathcal{M}}^4_p$ by blowing up the inf\/inite base of all light cones
in $\mathbb{P}^3(\mathbb{R})= \mathbb{P}^4 (\mathbb{R})\setminus {M}^4$.

\subsection{A warning: the orientation and topology of time}

For our present purposes, any time-like line in $\mathcal{M}^4$ may serve as a~substitute of ``time axis''.
It has the topology of the Euclidean line $\mathbb{R}$.

In both compactif\/ications that we have considered, $\overline{\mathcal{M}}^4_p$ and $\overline{\mathcal{M}}^4_q$, any
time-like line has the topology of a~{\it circle $S^1$}: it is completed by just one point lying on the inf\/inite
hyperplane $\mathbb{P}^3(\mathbb{R})$, resp.~$\mathbb{P}^4(\mathbb{R})$.
Even if we orient this time circle, its inf\/inite past coincides with its inf\/inite future.

Time in this picture can be imagined as moving along a~real projective line $\mathbb{P}^1(\mathbb{R})_{\rm time}$.
If we do not want to identify $0$ with~$\infty$, the beginning and the end of times, we must slightly change the
def\/inition of projective compactif\/ication of real space and the def\/inition of real blow-up.
This is done in Section~\ref{Section3} below.
Brief\/ly, physical time is {\it oriented}, because it f\/lows irrevocably from past to future.
Hence if we imagine a~compactif\/ication of space-time compatible with the idea of physical time, it is natural to add to
each time-like line {\it two points}: its ``inf\/inite past'' $-\infty$ and ``inf\/inite future'' $+\infty$.
Mathematically, this leads to the consideration of the two-fold cover of the former $\mathbb{P}^1(\mathbb{R})_{\rm time}$.
This cover topologically is still $S^1$, and the two points in each f\/iber of the cover correspond to two possible time
orientations.
Real blow-ups are def\/ined similarly.

For the discussion below and in Section~\ref{Section4}, it will be important to allow also ``complex-valued time'' and therefore to
imagine $\mathbb{P}^1(\mathbb{R})_{\rm time}$ embedded into $\mathbb{P}^1(\mathbb{C})_{\rm time}$, the ``complex projective line
of time''. Using a~more sophisticated model of our Universe, namely, the Friedman--Robertson--Walker one, we will argue
that $\mathbb{P}^1(\mathbb{C})_{\rm time}$ naturally arises in it as the {\it modular curve} parametrizing elliptic curves.

In this case, we can imagine the time axis during one aeon modelled by the {\it positive real semi-axis} $[0,+\infty]$
in $\mathbb{P}^1(\mathbb{C})_{\rm time}$.
This picture will allow us to appeal to some quantum ideas related to {\em Wick's rotation}, when time becomes purely
imaginary.

\subsection{Penrose's cyclic cosmology}\label{Section1.3}

Roger Penrose (see~\cite{Pe1} and earlier publications) suggested that our observable Universe that started with the Big
Bang was preceded by another stage (``aeon'') of its development that ended as cold, inf\/inite space (predicted to be the
f\/inal stage of our Universe as well).

This idea seemingly implies the break of continuity between geometries of space-time during the transition between two
aeons.
The way Penrose suggested to overcome this break consisted in matching {\it not the metrics} of the respective
space-times but {\it the conformal classes} of these metrics: he argued that rescaling the relevant Einstein metrics~by
conformal factors tending to zero, resp.\
inf\/inity, one can avoid the apparent discrepancy: cf.~a brief summary at pp.~204--205 of~\cite{Pe1} and the f\/irst
paragraph of~\cite{Ne}.

We argue that Penrose's joining of two aeons can be modeled in our picture by identifying the future boundary of
a~previous aeon with the past (Big Bang) boundary of the next aeon, ``crossover geometry''.

As a~model of the crossover geometry between $\overline{\mathcal{M}}^4_p$ and $\overline{\mathcal{M}}^4_q$ in our
example, we suggest a~choice from two possibilities.

{\em Crossover model I.} Identify the 3-dimensional projective space ${\rm bl}_x^{-1}(x)$ with 3-dimensional projective space at inf\/inity of
$\mathbb{P}^4 (\mathbb{R})= \overline{\mathcal{M}}^4_p$ in such a~way that the sphere of null-directions in
${\rm bl}_x^{-1}(x)$ is identif\/ied with the common base at inf\/inity of all light cones in ${\mathcal{M}}^4$.

The space-time containing two aeons will consist of two irreducible components intersecting along the common crossover
boundary $\mathbb{P}^3(\mathbb{R})$.

{\em Crossover model II.} In this model, the blowing up of the inf\/inite $S^2$ is a~model $C_{-}$ of the f\/irst aeon
preparing itself for the next Big Bang.
The blowing up of~$x$ is a~model $C_+$ of a~Big Bang of the second aeon.
Finally, the diagram~\eqref{(1.1)} describes the geometry of matching the two aeons: it shows that the divisor
${\rm bl}_{S^2}^{-1}(\mathbb{P}^3(\mathbb{R})_{\infty})\subset C_{-}$, ``inf\/inity of the f\/irst aeon'', can be identif\/ied with
the divisor ${\rm bl}_x^{-1}(L_x)\subset C_+$, the Big Bang of the next aeon, and after such an identif\/ication we get
a~connected space, say $C_{-}*C_+$, that can serve as a~geometric model of the space-time including both aeons.

The f\/irst crossover model is simpler and looks more universal.
The choice between the two possibilities may be a~matter of comparison with the geometry of the adopted dif\/ferential
geometric picture of the respective Einstein universes.
There the structure of boundary is dictated by the considerations similar to those that led to the understanding of the
Mixmaster model, see Section~\ref{Section4} below.

\subsection{Plan of the paper}
\label{1.4}

The next Section~\ref{Section2} is dedicated to the complexif\/ied space-times, involving spinors and Penrose's twistors.
This context is convenient for introducing conformal classes of metrics in one framework with all necessary geometric
tools.
Moreover, basic physical f\/ields, Lagrangians, and equations of motion (preceding quantisation) become very natural
constructions: cf.~Appendices in~\cite{Pe1} and the survey~\cite{Ma1}.
In particular, we consider a~complex version of the Big Bang diagram~\eqref{(1.1)}.

In the Section~\ref{Section3} we return to real models of space-time and discuss oriented versions of the diagram~\eqref{(1.1)} and
the respective notions of boundaries that arise in real algebraic geometry.
They should be compared with more physical treatments: see in particular survey~\cite{FlHeSa} and references therein.
These considerations refer to what can be called ``kinematic of boundaries''.

Section~\ref{Section4} introduces an element of dynamic, namely the picture of time ``on and near the boundary'', or during and
around the crossover.

After discussing the notion of cosmological time(s), the scheme we suggest is a~formalisation of the intuitive idea that
on the boundary ``at the moment of Big Bang'' time is purely imaginary.
The complex projective line of physical time mathematically appears as the modular curve parametrizing the family of
elliptic curves appearing in the description of Friedman--Robertson--Walker model, see Section~\ref{4.2} below.
We suggest that the inverse Wick rotation needed to make time real is mediated by the evolution along a~stretch of
hyperbolic geodesic on the upper (or rather {\it right}, see Section~\ref{4.5}) complex half-plane~$H$ which is the
standard cover of the respective modular curve.
This allows us to include into our picture the chaos of early Mixmaster Universe, whose standard description involves
exactly the same symbolic dynamics as that of hyperbolic geodesics, see Sections~\ref{4.3}--\ref{4.5}.

Notice that if we measure the cosmological time after Big Bang in terms of inverse temperature of the background
radiation $1/kT$, the backward passage to the imaginary time $it/h$ generally transforms various partition functions of
the chaos into their quantum versions, traces of evolving quantum operators.

If we want to use this picture for a~description of crossover between two aeons, it remains only to assume that the real
time of the previous aeon becomes imaginary at its future boundary.

We feel that here even our drastically simplif\/ied models must include elements of comparison with quantisation in order
to describe what is going near the common boundary of two aeons.
Traditionally, suf\/f\/iciently symmetric models of space-time are quantized as Hamiltonian systems of classical mechanics
(ADM quantisation).
For a~treatment of Bianchi models and Mixmaster solutions in this way see, e.g.,~\cite{Fu} and~\cite{YaHa}.
For the relation to
modular curves and modular symbols see~\cite{MaMar1,MaMar2}.

In the general algebraic spirit of this paper, the last Section~\ref{Section5} discusses options in the framework of
$C^*$-quantisation formalisms.
See also~\cite{EsMar,GrMaTe} and references therein.

\section{A complex Big Bang model}\label{Section2}

\subsection{Twistors and Grassmannian spinors}

The complex version of $\overline{\mathcal{M}}^4_p$ we consider is simply $\mathbb{P}^4$ (or rather
$\mathbb{P}^4(\mathbb{C})$, but we will sometimes omit mentioning $\mathbb{C}$-points explicitly).

The complex version of $\overline{\mathcal{M}}^4_q$ is the Grassmannian ${\rm Gr}(2,T)$ of 2-dimensional subspaces in the
4-dimensional complex vector space~$T$, {\it Penrose's twistor space}.

Below we brief\/ly recall the relevant geometric data.
For more details, see~\cite[Chapter~1]{Ma1}, in particular Section~3.

This Grassmannian carries the tautological vector bundle~$S$ whose f\/iber over a~point~$x$ is the subspace $S(x)\subset
T$ corresponding to this point.
Moreover, we need the second spinor bundle $\widetilde{S}$, whose f\/iber $\widetilde{S}(x)$ over~$x$ is the subspace
orthogonal to~$S(x)$ in the dual twistor space~$T^*$.
We use sometimes the respective sheaves of sections denoted $\widetilde{S}$ and $\widetilde{S}^*$ respectively.

The Grassmannian ${\rm Gr}(2,T)$ is canonically embedded into the projective space of lines in $\Lambda^2T$,
$P(\Lambda^2(T))\cong \mathbb{P}^5$, by the map $S\mapsto \Lambda^2(S)$ for each 2-dimensional subspace $S\subset T$.
The image of this embedding~$p$ is a~4-dimensional quadric hypersurface~$G$ \cite[Chapter~1, Section~3.2]{Ma1}.
We have canonical isomorphisms
\begin{gather}
\label{(2.1)}
\Lambda^2(S)=p^*(O_{P^5}(-1)),
\qquad
\Lambda^2(\widetilde{S})=\Lambda^2(S)\otimes \Lambda^4(T)
\end{gather}
(see~\cite[Chapter~1, Section~1.4]{Ma1}).

\subsection{Tangent/cotangent bundles and light cones}

Moreover, we have canonical isomorphisms~\cite[Chapter~1, Section~1.6]{Ma1}
\begin{gather}
\label{(2.2)}
\mathcal{T}_G=S^*\otimes \widetilde{S}^*,
\qquad
\Omega^1_G=S\otimes \widetilde{S}.
\end{gather}
``Null'' or ``light'' tangent vectors at a~point~$x$, by def\/inition, correspond to the decomposable tangent directions
$s^*(x)\otimes\tilde{s}^*(x)$ where $s^*(x)\in S^*(x)$, $\tilde{s}^*(x)\in \widetilde{S}^*(x)$.
A~line in~$G$ all of whose tangent vectors are null-vectors is called a~light ray.
Let $\mathbb{P}^4_x$ be the hyperplane in $\mathbb{P}(\Lambda^2(T))$ tangent to~$G$ at a~point~$x$.
Then $\mathbb{P}^4_x\cap G$ is a~singular quadric, the union of all light rays passing through~$x$ in~$G$, that is,
complex light cone with vertex~$x$.
In view of~\eqref{(2.2)}, the base of this cone is canonically identif\/ied with $\mathbb{P}^1(S^*)\times
\mathbb{P}^1(\widetilde{S}^*)$ \cite[Chapter~1, Section~3.6]{Ma1}.

\subsection{Conformal metrics}

In this context, it is natural to def\/ine a~conformal (class of) metric(s) as an invertible subsheaf of $S^2(\Omega^1)$
which is locally a~direct summand.
From~\eqref{(2.2)} one sees that on~$G$, we have such a~subsheaf $\Lambda^2(S)\otimes \Lambda^2(\widetilde{S})$.
A~choice of the local section of $\Lambda^2(S)\otimes \Lambda^2(\widetilde{S})$ determines an actual (complex) metric
wherever this section does not vanish.

Since $\Lambda^2(S)\otimes \Lambda^2(\widetilde{S})\cong p^*(O_{P^5}(-2))$ (see~\eqref{(2.1)}), any such metric must
have a~pole on the intersection of~$G$ with some quadric (possibly reducible, or even a~double hyperplane) in
$\mathbb{P}^5$, for example, a~union of two light cones, that may even coincide.
If we wish to compensate for this pole, we must locally multiply the metric by a~meromorphic function vanishing at the
polar locus.
Imagining this polar locus as a~``time horizon'' of the Universe, we are thus bridging our picture with (real) conformal
constructions by Penrose et al.

Finally, since for the def\/inition and study of curved complex space-times the basic structure consists precisely of
postulating two spinor bundles and an isomorphism~\eqref{(2.2)} (see~\cite{Pe3} and~\cite[Chapter~2]{Ma1}), essential features of
the complexif\/ied blow up construction~\eqref{(1.1)} can be generalized as we do below.
We chose to describe only the local picture; it may become a~part of a~large spectrum of more global models.

\subsection{A complex blow up diagram}
\label{2.4}

Consider two complex four-dimensional manifolds $\overline{\mathcal{M}}_{-}$ and $\overline{\mathcal{M}}_{+}$, non
necessarily compact, with the following supplementary structures.

$(a)$ \textit{A~smooth complex projective two-dimensional quadric $\mathcal{S}_{-}\cong \mathbb{P}^1\times \mathbb{P}^1$
embedded as a~closed submanifold into $\overline{\mathcal{M}}_{-}$.}

$(b)$ \textit{A~three-dimensional complex space $\mathcal{L}$ isomorphic to a~neighborhood of the vertex of the complex cone
with base $\mathbb{P}^1\times \mathbb{P}^1$ embedded as a~closed submanifold into $\overline{\mathcal{M}}_{+}$.}

If we blow up the vertex $x\in \overline{\mathcal{M}}_{+}$, the divisor $\mathbb{P}^3$ that replaces this vertex will
contain the quadric $\mathcal{S}_+$ of null-directions.
Denote by $\widetilde{\mathcal{M}}_{+}$ the result of such a~blow up.

The last piece of the data we need is:

$(c)$ \textit{An explicit isomorphism of $\mathcal{S_{-}}$ in $\overline{\mathcal{M}}_{-}$ with the quadrics of null
directions $\mathcal{S}_+$ in $\overline{\mathcal{M}}_{+}$.}

These data will be (local, complex) analogs of $\overline{\mathcal{M}}^4_p$ endowed with heavens $S^2$, and of
$\overline{\mathcal{M}}^4_q$ endowed with the light cone $L_x$ respectively, described in the Introduction.

Our complexif\/ied model of the transition between two aeons is then the connected sum
$\overline{\mathcal{M}}_{-}*{}_{\mathcal{S}}\widetilde{\mathcal{M}}_{+}$ in which two complex quadrics $\mathcal{S}_{-}$
and $\mathcal{S}_+$ are identif\/ied.

\subsection{A multiverse model in twistor space}\label{2.5}

We consider here a~version of the blow-up and gluing construction carried out in twistor space, after composing with the
Penrose twistor transform, and we relate the resulting ``multiverse picture" to moduli spaces of conf\/igurations of trees
of projective spaces recently introduced and studied from an algebraic geometric perspective in~\cite{ChGiKr}.

The Penrose transform is the correspondence given by the f\/lag variety $F(1,2;T)$, with~$T$ a~complex $4$-dimensional
vector space, with projection maps to the complexif\/ied space-time given by the Grassmannian ${\rm Gr}(2;T)$ and to the twistor
space given by the complex projective space $\mathbb{P}^3(\mathbb{C})$,
\begin{gather}
\label{(2.3)}
\mathbb{P}^3={\rm Gr}(1;T) {\longleftarrow} F(1,2;T) {\longrightarrow} {\rm Gr}(2;T).
\end{gather}
We refer the reader to~\cite[Chapter 1, Section 4]{Ma1}, for a~more detailed exposition.
The Penrose diagram~\eqref{(2.3)} corresponds to the collection of~$\alpha$-planes in the Klein quadric ${\rm Gr}(2;T)
\hookrightarrow \mathbb{P}^5$.
These planes, in which every line is a~light ray, give one of the two families of planes corresponding to the two
$\mathbb{P}^1$'s in the base of a~light cone $C(x)$.
The planes in the~$\alpha$-family are given by the second projection of the f\/ibers of the f\/irst projection
in~\eqref{(2.3)}.
The other family of planes, the~$\beta$-planes, are obtained similarly from the dual Penrose diagram
\begin{gather*}
{\rm Gr}(3;T^*) {\longleftarrow} F(2,3;T^*) {\longrightarrow} {\rm Gr}(2;T).
\end{gather*}

\subsection{Pointed rooted trees of projective spaces}

We consider oriented rooted trees that are f\/inite trees with one outgoing f\/lag (half-edge) at the root vertex and
a~number of incoming half-edges, with the tree oriented from the inputs to the root.
Each vertex~$v$ in the oriented rooted tree has one outgoing f\/lag and ${\rm val}(v)-1$ incoming f\/lags, each oriented edge~$e$
in the tree is obtained as the matching of the unique outgoing f\/lag at the source vertex with one of the incoming f\/lags
at the target vertex.

A rooted tree of projective spaces $\mathbb{P}^d$ is a~rooted tree~$\tau$ as above with a~projective space
$X_v=\mathbb{P}^d$ assigned to each vertex~$v$, with a~choice of a~hyperplane $H_v \subset X_v$, and of a~point $p_{v,f}
\in X_v$ for each incoming f\/lag~$f$ at~$v$, so that $p_{v,f}\neq p_{v,f'}$ for $f\neq f'$ and $p_{v,f}\notin H_v$ for
all~$f$.
For each edge~$e$ in the rooted tree~$\tau$ we consider the blowup of the projective space $X_{t(e)}$ of the target
vertex at the point $p_{t(e),f_e}$, and we glue the exceptional divisor $E_{t(e)}$ of the blowup
${\rm bl}_{p_{t(e),f_e}}(X_{t(e)})$ to the hyperplane $H_{s(e)} \subset X_{s(e)}$.
When all these blowups and identif\/ications are carried out, for all edges of~$\tau$, one obtains a~variety $X_\tau$, is
a~pointed rooted tree of projective spaces.

We consider the case where $d=3$.
Each $\mathbb{P}^3$ in a~tree of pointed projective spaces can be though of as the twistor space of a~$4$-dimensional
complex spacetime.

A hyperplane~$H$ in the twistor space $\mathbb{P}^3$ corresponds, under the Penrose transform, to a~$\beta$-plane in the
Klein quadric, and points in $\mathbb{P}^3$ correspond to~$\alpha$-planes.
Thus, the choice of a~hyperplane $H_v \subset X_v$ corresponds to f\/ixing a~$\beta$-plane in each copy $Q_v$ of the Klein
quadric, while the choice of distinct points $p_{v,f}$ in $X_v$ not on the hyperplanes corresponds to a~choice
of~$\alpha$-planes that do not meet the chosen~$\beta$-plane.

We then come to a~dif\/ferent model of gluing than the one discussed earlier, where the gluing is performed by blowing up
the twistor spaces~$X_v$ at the marked points $p_{v,f_e}$ of incoming f\/lags~$f_e$ corresponding to oriented edges~$e$
of~$\tau$ with $v=t(e)$ and gluing the exceptional divisor of the blowup to the hyperplane $H_{s(e)}$ in the twistor
space $X_{s(e)}$.

This sequence of blowups and gluings produces a~variety $X_\tau$, which is not necessarily itself the twistor space of
a~smooth $4$-dimensional space-time.
However, one can proceed in a~way similar to the method used in~\cite{DoFr}, where one considers a~gluing of blowups of
twistor spaces and then deforms it to a~new smooth twistor space.
In our setting, the variety $X_\tau$ def\/ines a~point in the moduli space $T_{3,n}$ of stable deformations~$n$-pointed
rooted trees of projective spaces of~\cite{ChGiKr}, see also~\cite{MaMar3}.
A~path in $T_{3,n}$ from this point to a~point in the open stratum provides a~deformation to a~single smooth twistor
space with marked points and a~marked hyperplane.

We can therefore interpret the moduli space $T_{3,n}$ as a~multiverse landscape and any natural class of functions on
this moduli space as multiverse f\/ields.

\section{Real models and orientation}\label{Section3}

\subsection{Projective spaces of real oriented lines}

As we mentioned, the diagram~\eqref{(1.1)} is obtained by constructing f\/irst the respective diagram of algebraic
geometric blow-ups and then restricting it to real points (in an appropriate real structure).
In this section, we will discuss more physical versions of~\eqref{(1.1)} f\/irst, by passing to certain unfamif\/ied
coverings of the involved manifolds and second, by introducing ``cuts'' of these coverings that may be compared to
various boundaries considered by physicists: causal, conformal etc.
(see~\cite{FlHeSa, GeKrPe} and references therein).

We start with real projective spaces.

Since $\mathbb{P}^1(\mathbb{R})$ is topologically $S^1$, we have $\pi_1(\mathbb{P}^1(\mathbb{R}))\cong \mathbb{Z}$.
However, for $n\ge 2$ we have $\pi_1(\mathbb{P}^n(\mathbb{R}))\cong \mathbb{Z}_2$, and the universal covering of
$\mathbb{P}^n(\mathbb{R})$ topologically is a~certain double cover $S^n\to \mathbb{P}^n(\mathbb{R})$.

A more algebraic picture is this.
Having chosen real homogeneous coordinates in $\mathbb{P}^n$, we may identify $\mathbb{P}^n(\mathbb{R})$ with
$(\mathbb{R}^{n+1}\setminus \{0\})/ \mathbb{R}^*$ where the multiplicative group of reals $\mathbb{R}^*$ acts~by
multiplying all homogeneous coordinates of a~point by the same factor.
Now put
\begin{gather*}
\mathbb{P}^n_{\rm or}(\mathbb{R}):= \big(\mathbb{R}^{n+1}\setminus \{0\}\big)/ \mathbb{R}_+^*,
\end{gather*}
where $\mathbb{R}^*_+$ is the subgroup of positive reals.

Then the tautological map $\mathbb{P}^n_{\rm or}(\mathbb{R}) \to \mathbb{P}^n(\mathbb{R})$ is the universal cover for $n\ge
2$.
However, we may and will use this map also for $n=1$ and even $n=0$ since $\mathbb{P}^n_{\rm or}(\mathbb{R})$ obviously
parametrizes oriented lines in $\mathbb{R}^{n+1}$ for all values of $n \ge 0$.

\subsection{Boundaries}

In the situation of the previous subsection, consider the complete f\/lag in $\mathbb{R}^{n+1}$:
$\{0\}=\mathbb{R}^0\subset \mathbb{R}^1\subset \mathbb{R}^2\subset \dots \subset \mathbb{R}^{n+1}$, where $\mathbb{R}^m$
is the span of the f\/irst~$m$ vectors of the chosen basis.

Then we get a~chain of embeddings $\mathbb{P}^0_{\rm or}(\mathbb{R})\subset \mathbb{P}^1_{\rm or}(\mathbb{R})\subset \dots
\subset \mathbb{P}^n_{\rm or}(\mathbb{R})$.
Each $\mathbb{P}^m_{\rm or}(\mathbb{R})$ is embedded in the next $\mathbb{P}^{m+1}_{\rm or}(\mathbb{R})$ as~$m$-dimensional
equator $S^m$ of a~sphere~$S^{m+1}$.
This equator cuts $\mathbb{P}^{m+1}_{\rm or}(\mathbb{R})$ into two open subsets, say, $\mathbb{P}^{m+1}_{\pm}(\mathbb{R})$,
each of which embeds as a~big cell into~$\mathbb{P}^{m+1}(\mathbb{R})$; the choice of one of the components corresponds
to the choice of orientation.
The same reasoning applies to the equator itself $\mathbb{P}^m_{\rm or}(\mathbb{R})$; we may choose an open half of it, say
$\mathbb{P}^{m}_{+}(\mathbb{R})$ and add it to $\mathbb{P}^{m}_{+}(\mathbb{R})$, etc.
The resulting partial compactif\/ication of $\mathbb{P}^{n+1}_{+}(\mathbb{R})$ may be considered as a~compactif\/ication of
space-time by a~``future'' boundary.

In particular, if we identify the big cell in $\mathbb{P}^4(\mathbb{R})$ with Minkowski space, and the last homogeneous
coordinate with oriented time coordinate, we can choose the future part of the equa\-to\-rial~$\mathbb{P}^3_{\rm or}(\mathbb{R})$ as one where the future part of (any) light cone f\/inally lands.

Lifting the diagram~\eqref{(1.1)} in this way to the diagram of parts of oriented Grassmannians, we f\/inally get the
algebraic geometric picture ref\/lecting time orientation.

\subsection{Grassmannians of real oriented subspaces}

Similarly, the Grassmannian of real oriented~$d$-dimensional subspaces in $\mathbb{R}^{d+c}$ is the double cover of the
space of real points of the relevant complex Grassmanian, and it is the universal cover, if $cd\ge 2$.

One can extend the previous treatment of the case $d=1$ using ``matrix homogeneous coordinates'' on Grassmannians as
in~\cite[Chapter~1, Section~1.3]{Ma1}.
We will omit the details.

\subsection{Real points of a~complex Big Bang model}

Real structures of complex spaces endowed with spinor bundles and isomorphisms~\eqref{(2.2)} are discussed
in~\cite{Pe2,Pe3, Pe1} and~\cite{Ma1}.
In the local context of Section~\ref{2.4}
the relevant spaces of real points of the quadrics $\mathcal{S}_{\pm}$ must have topology of $S^2$.
One can also imagine the identif\/ication of these two $S^2$'s as projection of the cylinder $S^2\times \mathbb{R}$
smashing the light-like axis $\mathbb{R}$.
In this way the transition phase between two aeons is modeled by a~trip along all light lines starting at the boundary.
As we argue in the next section, physical time along a~light geodesic does not ``stop'' as is usually postulated, but
takes purely imaginary values.
This is an additional argument to try the same picture for the crossover time between aeons.

\section{Big Bang models and families of elliptic curves}\label{Section4}

\subsection{Time in cosmology and modular curves}
\label{4.1}

The primary notion of time in relativistic models is local: basically, along each time-like oriented geodesic the
dif\/ferential of its time function $dt$ is $ds$ restricted to this geodesic, where $ds^2$ is the relevant Einstein
metric.
Formally applying this prescription, we have to recognise that even in a~f\/lat space-time, along space-like geodesics
time becomes purely imaginary, whereas light-like geodesics, along which time ``stays still'', form a~wall.
The respective wall-crossing in the space of geodesics produces the Wick rotation of time, from real axis to the pure
imaginary axis.
Along any light-like geodesic, ``real'' time stops, however ``pure imaginary time f\/low'' makes perfect sense appearing,
e.g., as a~variable in wave-functions of photons.

Below we will describe a~model in which time is imaginary at the past boundary of the universe (or future boundary of
the previous aeon), but the reverse Wick rotation does not happen instantly.
Instead, it includes the movement of time along a~curve in the complex plane.
However, an important feature of this picture is that local times are replaced by a~version of cosmological,
i.e.\ global time, say $\Theta_{-}$, resp.\
$\Theta_+$, for the previous, resp.\
next aeon.

A good example of observable global time is the inverse temperature $1/kT$ of the cosmic microwave background (CMB) radiation.
It is accepted that the current value of it measures the global age of our Universe starting from the time when it
stopped to be opaque for light, about $38\cdot 10^4$ years after the Big Bang.

Another version involves measuring the redshift of observable galaxies, thus putting their current appearance on various
cosmological time sections of our Universe, so that the scientif\/ic picture of the observable Universe bears an uncanny
resemblance to Marcel Duchamp's classics of modernism {\em ``Nude descending a~staircase''.}

As we brief\/ly described in Section~\ref{1.4},
for us more important is not a~choice of a~concrete parametrisation of physical
time (although we will use it later) but the image of an oriented time curve in the compactif\/ied complex plane
$\mathbb{P}^1(\mathbb{C})_{\rm time}$.
On the mathematical side, $\mathbb{P}^1(\mathbb{C})_{\rm time}$ will appear as a~{\it modular curve} parametrizing elliptic
curves) that emerge, e.g.,~in Robertson--Walker and Bianchi models of the previous aeon.

Another $\mathbb{P}^1(\mathbb{C})$, a~``modular one'', contains the complex half-plane~$H$ upon which the modular group
${\rm PSL}(2,\mathbb{Z})$ acts.
We use an identif\/ication of $\mathbb{P}^1(\mathbb{C})_{\rm time}$ with $\Gamma\setminus H$ where~$\Gamma$ is
${\rm PSL}(2,\mathbb{Z})$ or a~f\/inite index subgroup of it.

We will now imagine cosmological times $\Theta_{\pm}$ as certain coordinate functions along the time curve {\em lifted
to the modular plane.} When the universe attains along the real axis the wall from the side of the previous aeon,
$\Theta_{-}=\infty$, the time curve moves to the imaginary axis containing the same point $\Theta_{-}=\infty=i\infty$,
and follows it, say, from $i\infty$ to $i\epsilon_{-}$.

The imaginary moment $i\epsilon_{-}$ is the beginning of the Big Bang of the next aeon.

After wall-crossing, time moves along a~hyperbolic geodesic in the direction of the real axis.
Along this geodesic, time has non-trivial real and imaginary components.
When it reaches the real axis at a~point $\epsilon_{+}$, it becomes real time $\Theta_+$ of the next aeon.

In fact, in this situation we should think about geodesics on the {\it right} complex half-plane of time: $-iH= \{z\in
\mathbb{C} |{\rm Re}z >0\}$: see Section~\ref{4.5} below.

Physical evolution of the universe along the stretch of the geodesic is in principle a~quantum phenomenon, unlike the
classical models of cosmic space-time that we use in order to describe the aeons outside of the transition region.

However, this idea of non-trivial reverse Wick rotation allows us to incorporate the picture of the Mixmaster (Bianchi
IX) Universe as a~statistical dynamics approximation to an unknown quantum f\/ields (or strings) picture of the Big Bang.
Moreover, in our context it appears to be compatible with the Penrose picture, although many mathematical details are
still to be worked out.

\subsection{Friedman--Robertson--Walker (FRW) universe and elliptic curves}
\label{4.2}

Following~\cite{To} and~\cite{Ne}, we describe a~previous aeon universe as (the late stage) of the FRW model.

In this model, the space-time (during one aeon) can be represented as the direct product of a~global time~$t$-axis and
a~maximally symmetric three-dimensional space section with a~metric of constant curvature~$k$.
We choose also a~f\/ixed time-like geodesic (``observer's history'') along which the metric is $dt^2$, and coordinatize
each space section at the time~$t$ by the invariant distance~$r$ from the observer and two natural angle coordinates
$\theta$, $\phi$ on the sphere of radius~$r$.
By rescaling the radial coordinate, we may and will assume that the curvature constant~$k$ takes one of three values:
$k=\pm 1$ or $0$.

This rescaling produces the natural unit of length, when $k\ne 0$, and the respective unit of time is always chosen so
that the speed of light is $c=1$.

The RW metric of signature (1,3) is then given by the formula
\begin{gather}
\label{(4.1)}
ds^2:= dt^2 - R(t)^2 \left[\frac{dr^2}{1-kr^2}+r^2\big(d\theta^2+\sin^2\theta d\phi^2\big)\right].
\end{gather}

It might be convenient to replace~$r$ in~\eqref{(4.1)} by the third dimensionless ``angle'' coordinate $\chi:= r/R(t)$.
Then~\eqref{(4.1)} becomes
\begin{gather*}
ds^2:= dt^2 - R(t)^2 \big[d\chi^2+S_k^2(\chi)\big(d\theta^2+\sin^2\theta d\phi^2\big)\big],
\end{gather*}
where $S_k(\chi) =\sin \chi$ for $k=1$, $\chi$ for $k=0$, and $\sinh \chi$ for $k=-1$.

Dynamics in this model is described by one real function $R(t)$: it increases from zero at the Big Bang of one aeon to
inf\/inity during this aeon which, after the imaginary axis/geodesic transition described above, becomes ``almost zero
time'' of the next aeon.

We scale $R(t)$ by putting $R=1$ ``now'', as in~\cite{To}.
Notations in~\cite{To} slightly dif\/fer from ours.
In his formula for metric (2),~$r$ is our~$\chi$, and $f_k(r)$ is our $S_k(\chi)$.

This function is constrained by the Einstein--Friedman equations (here with cosmological constant $\Lambda =3$), which
leads to the introduction of the elliptic curve given by the equation in the $(Y,R)$-plane
\begin{gather}
\label{(4.3)}
Y^2= R^4+aR+b
\end{gather}
(see~\cite[equation~(3)]{To} and~\cite[equation~(9)]{Ne}, where their~$S$ is the same as our~$R$).

Besides the proper time~$t$, and the scale factor $R(t)$, global time may be measured by its conformal version~$\tau$,
which according to~\cite[formula~(3)]{To} may be given as the integral along a~real curve on the elliptic
curve~\eqref{(4.3)}:
\begin{gather*}
\tau \cong \int_0^{R(t)} \frac{dR}{Y}.
\end{gather*}
A~physical interpretation of the coef\/f\/icients $a$, $b$ as characterising matter and radiation sources in~\eqref{(4.3)}, for
which we refer the reader to~\cite{OlPe} and~\cite{To}, shows that in principle {\em $a$, $b$ also depend on time},
although for asymptotic estimates, their values are usually f\/ixed by current observations.

We close this subsection by the following qualitative summary:

{\em In the FRW universe, the time evolution is essentially described by a~real curve on an algebraic
surface~\eqref{(4.3)} which is a~family of elliptic curves.}

Universal families of elliptic curves are parametrized by modular curves, and in the next subsection we will see
a~family of elliptic curves naturally emerges in the description of a~late stage of evolution of the FRW model.
In a~pure mathematical context, the reader is invited to compare our suggestion with the treatment of the Painlev\'e VI
equation in~\cite{Ma2} and the whole hierarchy of Painlev\'e equations in~\cite{Ta}.

\subsection{Bianchi IX universe and the modular curve}
\label{4.3}

As a~model of the universe of the next aeon emerging after the Big Bang we take here the Bianchi~IX space-time,
admitting ${\rm SO}(3)$-symmetry of its space-like sections.
Its metric in appropriate coordinates takes the following form:
\begin{gather}
\label{(4.5)}
ds^2=dt^2 -a(t)^2dx^2-b(t)^2dy^2-c(t)^2dz^2,
\end{gather}
where the coef\/f\/icients $a(t)$, $b(t)$, $c(t)$ are called scale factors.

A family of such metrics satisfying Einstein equations is given by {\it Kasner solutions},
\begin{gather}
\label{(4.6)}
a(t)=t^{p_1},
\qquad
b(t)=t^{p_2},
\qquad
c(t)=t^{p_3}
\end{gather}
in which $p_i$ are points on the real algebraic curve{\samepage
\begin{gather}
\label{(4.7)}
\sum p_i=\sum p_i^2=1.
\end{gather}
These metrics become singular at $t=0$ which is the Big Bang moment.}

Around 1970, V.~Belinskii, I.M.~Khalatnikov, E.M.~Lifshitz and I.M.~Lifshitz argued that almost every solution of the
Einstein equations for~\eqref{(4.5)} {\em traced backwards in time $t\to +0$} can be approximately described~by
a~sequence of points~\eqref{(4.7)}: see~\cite{KLKhShSi} for a~later and more comprehensive study.
The~$n$-th point of this sequence begins the respective~$n$-th {\em Kasner era}, at the end of which a~jump to the next
point occurs, see below.

A mathematically more careful treatment of this discovery in~\cite{BoNo} has shown that this encoding is certainly
applicable to {\it another dynamical system} which is def\/ined on the boundary of a~certain compactif\/ication of the phase
space of this Bianchi IX model and in a~sense is its limit.

What makes this dynamical system remarkable in our context is that the construction involves {\it a~nontrivial real blow
up at the $t=0$}, see details in~\cite{Bo}.
The resulting boundary, that we suggest to identify with the wall between two aeons, is an attractor, it supports an
array of f\/ixed points and separatrices, and the jumps between separatrices which result from subtle instabilities
account for jumps between successive Kasner's regimes, corresponding to dif\/ferent points of~\eqref{(4.6)}.

In what sense this picture approximates the actual trajectories, is a~not quite trivial question: cf.\ the last three
paragraphs of~\cite[Section~2]{KLKhShSi}, where it is explained that among these trajectories there can exist
``anomalous'' cases when the description in terms of Kasner eras does not make sense, but that they are, in a~sense,
inf\/initely rare.
See also the recent critical discussion in~\cite{LuCh}.

The remaining part of this section is dedicated to three subject matters:

$(a)$ a~description of the BKLL encoding of trajectories of Bianchi IX boundary solutions by sequences of points
of~\eqref{(4.7)};

$(b)$ a~description of encoding of (most) geodesics with f\/inite ends on the complex upper half-plane~$H$ by a~version
of the continued fractions formalism and their projections to the modular curve ${\rm PSL}(2,\mathbb{Z})\setminus H$;

$(c)$ a~suggestion that the appropriate identif\/ication of these two descriptions corresponds to the identif\/ication of
two evolutions, involving imaginary/complex time on and around the wall between two aeons.
In fact both aeons then contribute mathematically comparable pictures of the time curve traced on a~family of elliptic
curves.

\subsection{BKLL encoding of Kasner eras}\label{4.4}

Consider a~``typical'' solution (trajectory)~$\gamma$ of the Einstein equations for~\eqref{(4.5)} as $t\to+0$. Introduce
the local logarithmic time~$\Omega$ along this trajectory with inverted orientation.
Its dif\/ferential is $d\Omega:=-\frac{dt}{abc}$, and the time itself is counted from an arbitrary but f\/ixed moment.
Then $\Omega \to +\infty$ approximately as $-\log t$ as $t\to +0$, and we have the following picture (perhaps strictly
applicable only to the boundary system referred to above, see~\cite{KLKhShSi} and~\cite{BoNo}).

$(i)$ As $\Omega \cong - \log t \to +\infty$, a~``typical'' solution~$\gamma$ of the Einstein equations determines
a~sequence of inf\/initely increasing moments $\Omega_0<\Omega_1< \dots <\Omega_n< \dots$ and a~sequence of irrational
real numbers $u_n\in (1, +\infty)$, $n=0,1,2,\dots$.

$(ii)$ The time semi-interval $[\Omega_n,\Omega_{n+1})$ is called the~$n$-th Kasner era (for the
trajectory~$\gamma$).
Within the~$n$-th era, the evolution of $a$, $b$, $c$ is approximately described by several consecutive Kasner's formulas.
Time intervals where scaling powers $(p_i)$ are (approximately) constant are called Kasner's cycles.

$(iii)$ The evolution in the~$n$-th era starts at time $\Omega_n$ with a~certain value $u=u_n>1$ which determines
respective scaling powers during the f\/irst cycle {\it in their growing order}
\begin{gather}
\label{(4.8)}
p_1=-\frac{u}{1+u+u^2},
\qquad
p_2=\frac{1+u}{1+u+u^2},
\qquad
p_3=\frac{u(1+u)}{1+u+u^2}.
\end{gather}
The next cycles inside the same era start with values $u=u_n-1, u_n-2, \dots$, and scaling po\-wers~\eqref{(4.8)} corresponding
to these numbers.

$(iv)$
After $k_n:=[u_n]$ cycles inside the current era, a~jump to the next era comes, with parameter
\begin{gather*}
u_{n+1}=\frac{1}{u_n-[u_n]}.
\end{gather*}

This means that the natural encoding of all $(u_n)$ together is obtained by considering an irrational number $x >1$
together with its continued fraction decomposition
\begin{gather*}
x= k_0+\frac{1}{k_1+\frac{1}{k_2+\cdots}}:= [k_0,k_1,k_2,\dots].
\end{gather*}
The time f\/low is modelled by the powers of the discrete shift
\begin{gather*}
[k_0,k_1,k_2,\dots] \mapsto [0, k_0,k_1,k_2,\dots],
\qquad
x\mapsto \frac{1}{x}- \left[\frac{1}{x}\right].
\end{gather*}
Put $x_n=[k_n,k_{n+1},\dots]$.

$(v)$ We compare the initial time-point $\Omega_{n+1}$ of the next era with $\Omega_n$ by introducing the additional
parameter $\delta_n$ via
\begin{gather*}
\Omega_{n+1}=(1+\delta_n k_n(u_n+1/x_n))\Omega_n.
\end{gather*}
Then the information about both sequences $(u,\Omega)$ simultaneously can be encoded by two numbers $(x,y)\in (0,1)^2$,
and the time f\/low can be modelled by powers of the shift of {\it two-sided} sequences of natural numbers
\begin{gather*}
[\dots, k_{-2}, k_{-1}, k_0,k_1,k_2,\dots]
\end{gather*}
or else
\begin{gather*}
(x, y) \mapsto \left(\frac{1}{x}- \left[\frac{1}{x}\right], \frac{1}{y+[1/x]} \right).
\end{gather*}
where $y=[0,k_0,k_{-1},k_{-2},\dots]$.

More precisely, if we put then $\eta_n=(1-\delta_n)/\delta_n$, $x_n=u_n-k_n$, we get the following recursion relation:
\begin{gather*}
\eta_{n+1}x_n=\frac{1}{k_n+\eta_nx_{n-1}}.
\end{gather*}

This means that in terms of the variables $(x_n,\, y_n:=\eta_{n+1} x_n)$ the transition to the next era is described by
the (almost everywhere) invertible operator acting upon $[0,1]\times [0,1]$,
\begin{gather}
\label{(4.10)}
\widetilde T: \  (x,y) \mapsto \left(\frac{1}{x} - \left[\frac{1}{x} \right], \frac{1}{y+[1/x]} \right),
\end{gather}
which is studied in~\cite{May} and~\cite{KLKhShSi}.

$(vi)$ The rearrangement of scaling factors $p_i^{(n)}(u)$ in the increasing order induces generally a~non-identical
permutation of the respective coef\/f\/icients.

Namely, as~$u$ diminishes by 1, the old permutation is multiplied by (12)(3) (see~\cite[formula~(2.3)]{KLKhShSi}). When
the era f\/inishes, the permutation (1)(23) occurs (this is~\cite[formula~(2.2)]{KLKhShSi}).

This means that during one era, the largest exponent decreases monotonically, and governs the same scale factor, $a$, $b$,
or~$c$ which we will call {\it the leading one.} Two other exponents oscillate between the remaining pair of scaling
coef\/f\/icients.
The number of oscillations is about $k_n:=[u_n]$.

In order to keep track of the sequence of the leading scale factors as well, we should consider orbits of
${\rm PGL}(2,\mathbb{Z})$ acting upon $\mathbb{P}^1(\mathbb{Q})\times \mathbb{P}$ where $\mathbb{P}$ can be na\-turally
identif\/ied with $\mathbb{P}^1(\mathbb{F}_2) = \{1,0,\infty\}$.
More precisely, the fractional linear transformation $u\mapsto 1/u$ that corresponds to the transition to the new era,
introduces the permutation (1)(23) of $\{1,0,\infty\}$, whereas the passage to a~new cycle within one era is described
by the transformation $u\mapsto u-1$ which induces the permutation (12)(3), see~\cite{MaMar1}.

\subsection{Symbolic dynamics of the geodesic f\/low on the modular surface}\label{4.5}

Miraculously, the same map~\eqref{(4.10)} describes an appropriate Poincar\'e return map for the geodesic f\/low on the
modular surface ${\mathbb M}$ which is either ${\rm PSL}(2,\mathbb{Z})\setminus H$, or $\Gamma_0(2)\setminus H$, if we wish to
take into account Kasner cycles within each era as at the end of Section~\ref{4.4}, $(iii)$ above.
The relevant Poincar\'e section is essentially the lift of the imaginary semi-axis of~$H$ to ${\mathbb M}$.
For more details, see~\cite{ChMay,EsMar, Se}.
In our context, an explanation of this coincidence is given by postulating the return of cosmological time to its real
values mediated by a~stretch of a~hyperbolic geodesic.

A warning is in order here: when we embed the real time~$\tau$ curve using an invariant of the elliptic
curve~\eqref{(4.3)} in the previous aeon, and a~real geodesic on ${\mathbb M}$ in the following aeon, during the
transition period we should interpret the upper half-plane~$H$, or rather its compactif\/ied version $H\cup
\mathbb{P}^1(\mathbb{R})$ as having the standard complex coordinate $z=-i\tau$.
Then the part of the imaginary half-axis of~$H$ between~$i$ and $i\infty$ projects onto a~real closed curve in ${\mathbb
M}$ which is now simply a~particular Poincar\'e section, a~device for encoding more interesting ``chaotic'' time geodesics.
On the contrary, the ``imaginary time axis'' on the ``wall'' between aeons invoked in Section~\ref{4.1} above becomes now
$\mathbb{P}^1(\mathbb{R})$, the real boundary of~$H$.

Of course, the action of ${\rm PSL}(2,\mathbb{Z})$ upon $\mathbb{P}^1(\mathbb{R})$ is topologically bad, and one can see in it
the basic source of stochasticity during the transition period.

\section[Conformal gluing and $C^*$-models]{Conformal gluing and $\boldsymbol{C^*}$-models}
\label{Section5}

\subsection{Twisted spectral triples and conformal factors}

The notion of {\it twisted spectral triples} was introduced by Connes and Moscovici (see~\cite{CoMo}) in order to extend
the spectral triple formalism of noncommutative Riemannian spin geometry to type~III cases that arise in the geometry of
foliations and in other settings (see~\cite{GrMaTe} for some cases related to number theory and to Schottky
uniformizations).

The prototype example discussed in~\cite{CoMo}, which is also the most relevant one for our present purposes, is coming
from the behaviour of the Dirac operator under conformal changes of the metric.
We review it here brief\/ly for later use.

Let~$M$ be a~compact Riemannian spin manifold.
It is well known that the Riemannian geo\-met\-ry of~$M$ can be reconstructed from its canonical spectral triple
$(A,H,D)=(C^\infty(M),L^2(M,S)$, $D)$, where~$D$ is the Dirac operator.
Thus, the notion of Riemannian spin geometry can be extended to the noncommutative setting, via spectral triples, where
an abstract data $(A,H,D)$ now consist of a~(possibly noncommutative) involutive algebra~$A$, a~representation of~$A$~by
bounded operators on a~Hilbert space~$H$, and a~(densely def\/ined) self-adjoint operator~$D$ on~$H$ with compact
resolvent, satisfying the compatibility condition: boundedness of all commutators $[D,a]$ with elements $a\in A$.

Let $(M,g)$ be a~compact~$n$-dimensional Riemannian spin manifold with Dirac operator~$D$.
Consider a~conformal change of the metric $g' = \Omega^2 g$, where we write $\Omega =e^{-2 h}$, for a~real valued
function $h\in C^\infty(M)$.
As was observed in~\cite{CoMo}, after identifying the Hilbert spaces of square integrable spinors via the map
of~\cite{BouGo} scaled by $e^{nh}$, the relevant Dirac operators become related by $D'=e^h D e^h$.

Moreover, according to~\cite{CoMo}, if the algebra $A_Y$ becomes noncommutative, the Dirac operator $e^h D_Y e^h$ no
longer has bounded commutators with elements of the algebra.
In this case, the correct notion that replaces the bounded commutator condition is the twisted version of~\cite{CoMo}.

Namely, for $D_Y'=e^h D_Y e^h$ one requires the twisted commutators
\begin{gather*}
D_Y' a-  \sigma(a) D_Y' = e^h \big[D_Y, e^h ae^{-h}\big] e^h
\end{gather*}
to be bounded for all $a\in A_Y$, where $\sigma(a):= e^{2h}ae^{-2h}$.
This replaces the ordinary notion of a~spectral triple with the notion of twisted spectral triple.

\subsection{Time evolution and conformal factors}

The expression above for the twisted commutator suggests that, in the case of a~noncommutative algebra $A_Y$, one can
consider a~time evolution determined by the conformal factor, with $h=h^*\in A_Y$ and $t\in \mathbb{R}$,
\begin{gather*}
\sigma_t(a)=e^{-it h}  ae^{it h}.
\end{gather*}

Thus, consider the case of a~Riemannian $4$-dimensional geometry that is locally a~cylinder $X=Y\times I$ with a~metric
$g_X = dt^2 + g_{Y,t}$, where $g_{Y,t} = \Omega^2(t) g_Y$, for a~f\/ixed $g_Y$ and $\Omega(t)=e^{-2 h t}$, for some f\/ixed
$h \in C^\infty(Y,\mathbb{R}^*_+)$.
If the algebra $C(Y)$ admits a~noncommutative deformation compatible with the metric, then the transformation $a\mapsto
e^hae^{-h}$ that arises from the twisted commutator with the conformally rescaled Dirac operator can be seen as the
ef\/fect of an evolution in imaginary time, under an analytic continuation to imaginary time of the time evolution def\/ined
above.

\subsection{Noncompact and Minkowskian geometries}

The setting of spectral triples (and by extension twisted spectral triples) can be generalized to the non-compact case,
as in~\cite{GGISV}, by replacing the compact resolvent condition for the Dirac operator by a~local condition, namely
requiring that $a (D-\lambda)^{-1}$ is compact, for some $a\in A$ and $\lambda\notin {\rm Spec}(D)$.
See~\cite[Section~3]{GGISV} for more details on the properties of non-compact spectral triples.

Accommodating Minkowskian geometries within the setting of spectral triples is a~more delicate issue, because the
Lorentzian Dirac operator is no longer self-adjoint and it is not an elliptic operator.
A~commonly used approach replaces Hilbert spaces with Krein spaces~\cite{Bog}.
The case of f\/lat Lorentzian cylinders over tori and their isospectral noncommutative deformations was treated with these
techniques in~\cite{vSuij}.
Noncommutative Minkowskian geometry and isospectral noncommutative deformations were also considered in a~number
theoretic setting in~\cite{Mar}.

\subsection{Noncommutativity from isospectral and toric deformations}

A general procedure to obtain noncommutative deformations of a~commutative algebra of functions, in a~way that preserves
the metric structure, is through the {\em isospectral deformations} of~\cite{CoLa}.
Assume that the compact Riemannian manifold~$Y$ is endowed with an action $\alpha: T^2 \to {\rm Isom}(Y, g_Y)$ of a~torus
$T^2=U(1)\times U(1)$ by isometries.
One obtains then a~noncommutative deformation $A_{Y,\theta}$ of the algebra of functions $A_Y=C(Y)$, depending on a~real
parameter~$\theta$, by the following procedure.
Given $f\in C(Y)$, in the representation on the Hilbert space $H_Y=L^2(Y,S_Y)$, one decomposes the operator $\pi(f)\in
B(H_Y)$ into weighted components according to the action of $T^2$, $\alpha_{\tau}(\pi(f_{n,m}))=e^{2\pi i
(n\tau_1+m\tau_2)} \pi(f_{n,m})$ The deformed product is then given on components by $f_{n,m}\star_\theta h_{k.r}=e^{\pi
i \theta (nr-mk)} f_{n,m} h_{k,r}$.
Geometrically, this corresponds to deforming the torus $T^2$ to noncommutative torus $T^2_\theta$, where one def\/ines the
algebra of the noncommutative torus as the twisted group algebra $C^*(\mathbb{Z}^2,\gamma)$ with cocycle
$\gamma((n,m),(k,r))=\exp(\pi i \theta (nr-mk))$.
The deformation $A_{Y,\theta}$ is isospectral, in the sense that the data $(H_Y,D_Y)$ of the spinor space and Dirac
operator, that determine the metric structure, remains undeformed.
Theta deformations, for~$\theta$ a~skew-symmetric matrix, can similarly be obtained for compact Riemannian manifolds~$Y$
with an isometric action of a~higher dimensional torus $T^k =U(1)^k$, which is similarly deformed to a~noncommutative
torus $T^k_\theta$.

The construction of such~$\theta$-deformations in the setting of noncommutative Riemannian geometry was extended to an
algebro-geometric setting in the work of Cirio, Landi and Szabo~\cite{CiLaSza2,CiLaSza1,CiLaSza3}, by replacing the real
noncommutative tori with algebraic noncommutative tori.
This leads to noncommutative deformations of projective spaces and other toric varieties, as well as deformations of
Grassmannians, via a~deformation of their Pl\"ucker coordinates.

\subsection{Noncommutative deformations of complex space-times}

Consider the complex $4$-manifolds $\overline{\mathcal{M}}_\pm$ involved in the blowup diagram giving the gluing of
successive aeons in our complex big bang model.
As we have seen, these are, respectively, the projective space $\mathbb{P}^4$ and the Grassmannian ${\rm Gr}(2,T)$ of
$2$-planes in a~$4$-dimensional complex vector space~$T$.
We view the Grassmannian ${\rm Gr}(2,T)$ as embedded in $\mathbb{P}(\Lambda^2 T)\simeq \mathbb{P}^5$ under the Pl\"ucker map,
with image the Klein quadric~$Q$ in $\mathbb{P}^5$.

The appropriate noncommutative~$\theta$-deformation for the Klein quadric (and for more general Grassmannians) was
constructed in~\cite{CiLaSza2,CiLaSza1}, in terms of homogeneous coordinates in~$\theta$-deformed projective
spaces and noncommutative Pl\"ucker relations.

More precisely, we have~$\theta$-deformations $\mathbb{P}^n_\theta$ of projective spaces, whose homogeneous coordinate
algebra has generators $\{w_i \}_{i=1,\ldots,n+1}$ and relations $w_i w_j = q_{ij}^2 w_j w_i$, for $i,j=1,\ldots, n$,
with $q_{ab}=\exp(\frac{i}{2} \theta_{ab})$ and $w_{n+1} w_i = w_i w_{n+1}$, for $i=1,\ldots, n$.

The algebra of functions on the noncommutative Grassmannian ${\rm Gr}_\theta(2,T)$ has six generators,
$\{\Lambda^I=\Lambda^{(ij)} \}_{1\leq i<j\leq 4}$, labeled by minors~$I$ of a~$2\times 4$-matrix.
In general, for a~Grassmannian ${\rm Gr}_\theta(d;n)$ these variables $\Lambda^J$, for minors $J=(j_1,\ldots,j_d)$, satisfy
relations
\begin{gather*}
\Lambda^{J} \Lambda^{J'} = \left(\prod\limits_{\alpha,\beta=1}^d q^2_{j_\alpha,j_\beta'}\right) \Lambda^{J'} \Lambda^J,
\end{gather*}
as in~\cite[equation~(1.26)]{CiLaSza3}.

The skew-symmetric matrix~$\Theta$ is related to~$\theta$ as in~\cite[equation~(1.28)]{CiLaSza3}
\begin{gather*}
\Theta^{J J'} = \sum\limits_{\alpha,\beta =1}^d \theta^{j_\alpha j_\beta'},
\end{gather*}
as a~necessary and suf\/f\/icient condition for the existence of Pl\"ucker embedding.

The noncommutative Pl\"ucker embedding of the deformed ${\rm Gr}_\theta(2,T)$ in $\mathbb{P}^5_\Theta$ is then determined~by
the relation
\begin{gather*}
q_{31} q_{32} q_{34} \Lambda^{(12)} \Lambda^{(34)} - q_{21} q_{23} q_{24} \Lambda^{(13)} \Lambda^{(24)} + q_{12} q_{13}
q_{14} \Lambda^{(23)} \Lambda^{(14)} =0,
\end{gather*}
where $q_{ij}$ were def\/ined above.

It is also shown in~\cite{CiLaSza3} that there is a~compatible real structure on the deformed Grassmannian
${\rm Gr}_\theta(2,T)$ and a~unique~$\theta$-deformation of the sphere $S^4$ that is compatible with a~noncommutative twistor
correspondence.
As shown in~\cite[Section 2.3]{CiLaSza3}, this corresponds to the involutive $\star$-algebra structure on the
noncommutative Klein quadric for which $q_{12}=q_{21}^{-1}=q$ and the other $q_{ij}=1$.

The standard construction of Schubert cells is also compatible with this quantization.

The Grassmannian ${\rm Gr}(2,T)$, which gives the complexif\/ied spacetime, has a~cell decomposition into six Schubert cells
$C_{(j_1,j_2)}$, with
\begin{gather*}
(j_1,j_2)\in \{(1,2), (1,3), (1,4), (2,3), (2,4), (3,4) \},
\end{gather*}
respectively of complex dimensions~$0$,~$1$,~$2$,~$2$,~$3$,~$4$.
They correspond to $2\times 4$-matrices in row echelon form, and consist of $2$-planes~$V$ that intersect the standard
f\/lag~$F$ with $\dim(V\cap F_{j_\ell})=\ell$.
In terms of the Pl\"ucker embedding ${\rm Gr}(2,T) \hookrightarrow \mathbb{P}^5$, if we write the def\/ining equation for
${\rm Gr}(2,T)$ in $\mathbb{P}^5$ as above, in the form
\begin{gather*}
\Lambda^{(12)} \Lambda^{(34)} - \Lambda^{(13)} \Lambda^{(24)} + \Lambda^{(23)} \Lambda^{(14)} =0,
\end{gather*}
then the Schubert varieties $X_{(j_1,j_2)}$ given by the closures of the Schubert cells,
$X_{(j_1,j_2)}=\overline{C_{(j_1,j_2)}}$, are given, respectively, by equations
\begin{gather*}
X_{(1,2)}=\big\{V \in {\rm Gr}(2,T) \,\big|\, \Lambda^{(13)}=\Lambda^{(14)}= \Lambda^{(23)}=\Lambda^{(24)}=\Lambda^{(34)}=0 \big\},
\\
X_{(1,3)}=\big\{V \in {\rm Gr}(2,T) \,\big|\, \Lambda^{(14)}= \Lambda^{(23)}=\Lambda^{(24)}=\Lambda^{(34)}=0 \big\},
\\
X_{(1,4)}=\big\{V \in {\rm Gr}(2,T) \,\big|\, \Lambda^{(23)}=\Lambda^{(24)}=\Lambda^{(34)}=0 \big\},
\\
X_{(2,3)}=\big\{V \in {\rm Gr}(2,T) \,\big|\, \Lambda^{(14)}=\Lambda^{(24)}=\Lambda^{(34)}=0 \big\},
\\
X_{(2,4)}=\big\{V \in {\rm Gr}(2,T) \,\big|\, \Lambda^{(34)}=0 \big\},
\end{gather*}
with $X_{(3,4)}={\rm Gr}(2,T)$.

\subsection{Quantization and gluing of aeons}

By the explicit description of Schubert cells and Schubert varieties that we recalled above, we see that the
quantization ${\rm Gr}_\theta(2,T)$, given by deforming the Pl\"ucker embedding to
\begin{gather*}
q_{31} q_{32} q_{34} \Lambda^{(12)} \Lambda^{(34)} - q_{21} q_{23} q_{24} \Lambda^{(13)} \Lambda^{(24)} + q_{12} q_{13}
q_{14} \Lambda^{(23)} \Lambda^{(14)} =0,
\end{gather*}
induces compatible quantizations of the Schubert varieties $X_{(j_1,j_2)}$.
In particular, for the closure~$X_{(2,4)}$ of the $3 $-dimensional cell, with the set of quantization parameters given
by the $q_{ab}$ of ${\rm Gr}_\theta(2,T)$ with $a\neq 3$.

When we view ${\rm Gr}(2,T)$ as complex spacetime, the big cell $U=C_{(3,4)}$ is the complexif\/ied Minkowski space and the
Schubert variety~$X_{(2,4)}$ can be identif\/ied with the boundary $C(\infty)$ given by light rays through inf\/inity,
see~\cite[Chapter~1, Section~3.9]{Ma1}.
The description of the cone~$C(\infty)$ as the locus of $V\in {\rm Gr}(2,T)$ with $\dim(V \cap F_2)\geq 1$, with respect to the
chosen f\/lag~$F$, corresponds to the usual description of the codimension one Schubert cycle in~${\rm Gr}(2,T)$.

Let us then consider again, in these terms, the two Crossover models from Section~\ref{Section1.3}.
To make the case of Crossover model~I compatible with the~$\theta$-deformations, it suf\/f\/ices to use a~$\theta$
deformation of the exceptional divisor $\mathbb{P}^3$ of the blowup of $Q^4$ where the deformation parameters match the
deformation parameters of the $\mathbb{P}^3$ at inf\/inity of $\mathbb{P}^4$.

We consider then the case of Crossover model~II.
It is based upon identif\/ication of the intersection $\overline{L}_x=Q^4\cap \mathbb{P}^4_x$ in $\mathbb{P}^5$ and
a~compactif\/ied light cone in $\overline{\mathcal{M}}^4_p$, which leads to the blowup diagram of the transition between
aeons.
We check that the picture still makes sense when we pass to compatible~$\theta$-deformations.

In the case of $\overline{\mathcal{M}}^4_q$, the Klein quadric $Q^4$ in $\mathbb{P}^5$, we use the quantization
of~\cite{CiLaSza1} described above, with the compatible quantization on the Schubert varieties $X_{(j_1,j_2)}$.
In the case of $\overline{\mathcal{M}}^4_p=\mathbb{P}^4(\mathbb{C})$ we consider a~ruled surface def\/ined by the equation
$x_0 x_3 + x_1 x_2=0$ inside the $\mathbb{P}^3$ def\/ined by setting $x_4=0$.
This can be seen as the intersection of the Klein quadric with the $\mathbb{P}^3$ given by $x_4=0$ and $x_5=0$ in
$\mathbb{P}^5$.
Thus, we can compatibly quantize the $\mathbb{P}^4(\mathbb{C})$ and the hyperplane $\mathbb{P}^3$ cut out by $x_4=0$ and
the locus $x_0 x_3 + x_1 x_2=0$ using the same quantization parameters of that we used for $\mathbb{P}^5$ and the Klein
quadric $Q^4$ in it.

\subsection{Conformal cyclic cosmology versus Mixmaster universe}

Both Penrose's conformally cyclic cosmology model and the older Mixmaster universe model provide models of the universe
undergoing a~series of cycles, or aeons.
However, the two models dif\/fer signif\/icantly in the way the cycling happens and in the physical properties that describe
the behavior near the singularity.
A~comparative analysis of Mixmaster type models and conformally cyclic cosmology can be found in Sections~2.4, 2.6 and~3.1 of~\cite{Pe1}.

The main dif\/ference between these two models lies in the fact that conformally cyclic cosmology requires a~conformally
f\/lat isotropic spacetime, while the spacetime of the Mixmaster universe is highly anisotropic.
Moreover, conformally cyclic cosmology requires a~vanishing or at least bounded Weyl curvature at the singularity, which
is related to a~very low entropy by Penrose's Weyl curvature hypothesis~\cite{Pe4},
while the Weyl curvature diverges widely in the Mixmaster case, with a~high entropy singularity.
Thus, one can consider the Mixmaster model as representing the generic, typical, conditions of the collapse, with high
entropy and chaotic dynamics, and the conformally cyclic cosmology as a~very special low entropy scenario.

The algebraic geometric setting we consider here can be adapted to accommodate both the conformally cyclic cosmology to
a~geometry and a~Mixmaster type model.
We have seen how to describe the conformal gluing of eons in terms of blowup diagrams.
We now consider a~similar picture for a~Mixmaster type cosmology.

Let us consider again the case of a~ruled quadric $\mathcal{S} \simeq \mathbb{P}^1 \times \mathbb{P}^1$ in a~hyperplane
$\mathbb{P}^3 \subset \mathbb{P}^4$ and the blowup ${\rm bl}_\mathcal{S}(\mathbb{P}^4)$.
The exceptional divisor~$E$ of the blowup is a~projective bundle over $\mathcal{S}$ with f\/iber
$\mathbb{P}^{\text{codim}(\mathcal{S})-1}=\mathbb{P}^1$.

In the case of conformally cyclic cosmology, we considered the real structure as in~\cite[Chapter~1, Sections 3.9--3.19]{Ma1},
where the real involution on $\mathcal{S}$ comes from exchanging the two spinor bundles~$S$ and $\tilde S$,
so that $S(\mathbb{R})=S^2$, the diagonal $\mathbb{P}^1(\mathbb{C})$, interpreted as the ``sky".
We now consider instead a~dif\/ferent real structure where the real involution is the standard conjugation on each of the
two copies of $\mathbb{P}^1$, which gives a~set of real points $\mathcal{S}(\mathbb{R})=S^1 \times S^1$, given by the
two $\mathbb{P}^1(\mathbb{R})$.
In this model, the real $E(\mathbb{R})$ has (locally) a~product structure as a~circle-bundle over $T^2=S^1\times S^1$.
The gluing of successive aeons now happens along this circle-bundle.

\looseness=-1
In this case we restrict to considering metrics that are compatible with the local product structure on the exceptional
divisor~$E$ in the blowup, so that the metric restricted to~$E$ has the diagonal form $h_1(z_1) + h_2(z_2) + h_3(z_3)$
where each $h_i$ is a~Riemannian metric on $\mathbb{P}^1$.
We identify the gluing region locally as a~one parameter family with f\/iber~$E$ over a~line (representing the
complexif\/ied time coordinate).
Then, instead of modifying the metric as in the conformally cyclic cosmology, by multiplying by a~single meromorphic
function vanishing at inf\/inity, we can now consider non-isotropic metrics adapted to the product structure, of the form
$a(z)^2 h_1(z_1) + b(z)^2 h_2(z_2) + c(z)^2 h_3(z_3)$, with~$z$ the complexif\/ied time coordinate and, in particular,
metrics where the factors $a(z)$, $b(z)$, $c(z)$ are of Kasner form $a(z)=z^{p_1}$, $b(z)=z^{p_2}$ and $c(z)=z^{p_3}$ with
$\sum\limits_i p_i=\sum\limits_i p_i^2=1$, where the exponents $p_i$ are chosen, in each Kasner era, according to the
prescription described in Sections~\ref{4.4} and~\ref{4.5},
when the complex time coordinate moves along geodesics in the modular
$\mathbb{P}^1$.
We'll see below that this embedding of Mixmaster type cosmologies inside the complex blowup diagram of cyclic cosmology
leads to a~compatible quantization picture.

\subsection{Comparing quantizations}

We can also compare the quantization by~$\theta$-deformation of the conformally cyclic cosmology model described above
with the quantization by~$\theta$-deformation of the Mixmaster universe described in~\cite{EsMar}.
To obtain a~$\theta$-deformation of the gluing region in the Mixmaster type model described above, we consider an
algebraic~$\theta$-deformation (in the sense of~\cite{CiLaSza2,CiLaSza1}) of the $\mathbb{P}^1$-bundle~$E$ over
$\mathcal{S}=\mathbb{P}^1\times \mathbb{P}^1$, which induces a~$C^*$-algebraic~$\theta$-deformation on the circle bundle
$E(\mathbb{R})$ over $T^2$, locally modelled on a~noncommutative $3$-torus $T^3_\theta$.
This is compatible with the~$\theta$-deformations of the Mixmaster universe cosmology considered in~\cite{EsMar} based
on noncommutative $3$-tori.
Notice that, if we start from a~$\theta$-deformation of the ambient complex spacetime $\mathbb{P}^4$ (or ${\rm Gr}(2,T)$) which
induces a~compatible~$\theta$-deformation of $\mathcal{S}$ (see the discussion above about the cells of the
Grassmannian), then a~compatible deformation of~$E$ will maintain the f\/iber direction as a~distinct (commutative)
direction, while if we only require the quantization to happen at (or near) the gluing, then more
general~$\theta$-deformations of~$E$ can be considered, analogous to the noncommutative $3$-tori of~\cite{EsMar}.
The fact that one of the three spatial directions is singled out is compatible with the fact that the Mixmaster models
have, in each Kasner era, a~distinguished spatial direction that dominates the expansion or contraction.

\subsection{Predictive aspects}

In this paper we are primarily concerned with the development of mathematical models for various aspects of cosmology,
based on the use of algebro-geometric techniques and noncommutative deformations.

At this stage of development of these models, it is premature to focus on predictions that may be detected in
observational cosmology.
However, we point out some directions in which one may be able to explore more concrete predictive aspects, though these
would f\/irst require further theoretical developments.

It is well known that Penrose's cyclic cosmology model ran into serious dif\/f\/iculties in the verif\/ication of predictions
on signatures of ripples from previous aeons in the background radiation.
In particular, the evidence proposed in~\cite{GuPe} was seriously criticized in~\cite{Haj,WeEr},
and~\cite{MoScZi}, see also the answer of~\cite{GuPe2}.
At the theoretical level, Penrose's model is also criticized for the lack of a~specif\/ic physical process that causes the
described transition between aeons (unlike the dynamical f\/ields responsible for a~similar behavior in the
Steinhardt--Turok cyclic universe).
As observed in~\cite{To}, a~bound on the angular size of the circles hypothesized by Penrose (as would follow
from~\cite{GuPe} and~\cite{GuPe2}) would not be compatible with inf\/lation in the early universe.

\looseness=-1
The theory of inf\/lationary cosmology recently received a~spectacular observational veri\-f\/i\-ca\-tion, along with primordial
gravitational waves, announced by the BICEP2 team~\cite{BICEP2}.
However, these f\/indings are also currently being scrutinized for possible cosmic dust contributions to the signal,
see~\cite{Zal}, awaiting an independent analysis based on the Planck data, presently still to be released.

The multiverse picture presented in Section~\ref{2.5}
f\/its more closely within the inf\/lationary models than Penrose's original
cyclic cosmology scenario.
In particular, it may be related to the type of eternal inf\/lation scenarios developed, in a~discretized model,
in~\cite{HSSS}.
Thus, possible predictions may be obtainable in this setting by developing a~more detailed analysis of the ``multiverse
f\/ields'' outlined in Section~\ref{2.5}, as in~\cite{HSSS}.
Algebro-geometric methods in the context of~\cite{HSSS} have already been considered in~\cite{MarTe}.

In a~dif\/ferent direction, one can seek predictions with possible observational verif\/ication, pertaining to the
quantization picture described in Section~\ref{Section5}.
This is based on noncommutative deformations of spacetime near the singularity.
There has been considerable work in the past on obtaining predictions of signals of early universe noncommutativity that
would be detectable in the CMB sky, see for instance~\cite{FuKoMi,Nau,SMRA}.
The question, in this case, would be to identify specif\/ic signatures of the type of noncommutative deformation
considered here, based on the $\theta$-deformation method.

\subsection*{Acknowledgements}

This paper was conceived after the lecture in Bonn (November 2013), in which Sir Roger Penrose explained his fascinating
ideas about cyclic cosmology.
Ya.~Sinai and O.~Bogoyavlenskii made helpful remarks about BKLL treatment of the Bianchi~IX model.
We are grateful to them.

\pdfbookmark[1]{References}{ref}
\LastPageEnding

\end{document}